\documentclass[preprint,10pt]{elsarticle}

\usepackage{graphicx}
\usepackage{amssymb}
\usepackage{amsmath}
\usepackage{bm}

\biboptions{square,sort&compress}
\journal{High Energy Density Physics}

\begin{document}

\begin{frontmatter}

\title{Resonant Bound-Free Contributions to Thomson Scattering of X-rays by Warm Dense Matter}

\author{W. R. Johnson}
\ead{johnson@nd.edu}
\address{University of Notre Dame, Notre Dame, IN 46556}
\author{J. Nilsen}
\ead{nilsen1@llnl.gov}
\author{K. T. Cheng}
\ead{ktcheng@llnl.gov}
\address{Lawrence Livermore National Laboratory, Livermore, CA 94551}

\begin{abstract}
Recent calculations [Nilsen et al.\ arXiv:1212.5972] predict that contributions
to the scattered photon spectrum from $3s$ and $3p$ bound states in chromium ($Z=24$) at
metallic density and $T=12$~eV
resonate below the respective bound-state thresholds.
These resonances are shown to be closely related to continuum lowering,
where $3d$ bound states
in the free atom dissolve into a resonant $l=2$ partial wave in the continuum.
The resulting $d$-state resonance
dominates contributions to the
bound-free dynamic structure function, leading to the predicted resonances
in the scattered X-ray spectrum.
Similar resonant features are shown to occur in all elements in the periodic table
between Ca and Mn  ($20\leq Z\leq 25$).
\end{abstract}

\begin{keyword}
52.25.Os: Emission, absorption and scattering of radiation \sep
52.38.-r: Laser-plasma interaction \sep
52.70.-m: Plasma diagnostic techniques \sep
56.65.Rr: Particle in cell method.
\end{keyword}
\end{frontmatter}


As discussed in the accompanying article Nilsen et al.\ \cite{ar:12}, resonances in bound-free contributions
to Thomson scattering of X-rays for a Cr plasma can complicate the determination of plasma density and
temperature from measurements of the scattered X-ray spectrum.
Inasmuch as bound-state features in dense plasmas of
lighter elements show up as smooth broad features, it is of interest to explore the origin of the resonances
that show up in Cr. To this end, we first note that the free Cr atom has the configuration
[Ar] $3d^5\,4s$. In a dense plasma, the bound electrons outside the Ar-like ionic core dissolve into
the continuum.
In particular, the $3d$ electrons in Cr form a resonant state near 12~eV above threshold
for the conditions comsidered here.
It is the contribution of this $d$-state  resonance that leads to resonances in the Thomson scattering
cross section. In the following paragraphs, we study these resonances in greater detail
for Cr and other elements between Ca and Mn ($20\leq Z \leq 25$).

\begin{table}
\caption{Properties of elements $20 \leq Z \leq 25$ at $T = 5$~eV. $A$ is the atomic weight,
$\rho$ (g/cc) is plasma density,
$\epsilon_{2s}$ and $\epsilon_{2p}$ are average-atom eigenvalues (eV) of $3p$ and $3s$ bound states,
$N_b$ and $N_c$ are the number of bound and
continuum electrons per ion. $Z_i$ is the average ion charge.
$n_s$, $n_p$ and $n_d$ are the number of continuum electrons
in $s$, $p$ and $d$ states, respectively. \label{tab1}}
\begin{center}
\begin{tabular}{|crrrrrr|}
\hline\hline
       &    Ca   &  Sc    &   Ti   &   V    &   Cr   &   Mn   \\
       \hline
 $Z$   &    20   &   21   &   22   &   23   &   24   &   25   \\
 $A$   &   40.080&  44.960&  47.870&  50.940&  52.000&  51.940\\
 $\rho$&    1.56 &    2.99&    4.54&    6.11&    7.19&   7.47 \\[1ex]
 -$\epsilon_{3s}$ &    35.16&   38.94&   43.68&   49.50&   56.22&   63.66\\
 -$\epsilon_{3p}$ &    16.99&   18.69&   21.26&   24.84&   29.26&   34.37\\[1ex]
 $N_b$ &    17.67&   17.91&   17.96&   17.97&   17.98&   17.99\\
 $N_c$ &     2.33&    3.09&    4.04&    5.03&    6.02&    7.01\\
 $Z_i$ &     1.45&    2.00&    2.24&    2.56&    2.65&    2.32\\[1ex]
 $n_s$ &    0.298&   0.389&   0.419&   0.474&   0.511&   0.507\\
 $n_p$ &    0.459&   0.572&   0.606&   0.691&   0.744&   0.709\\
 $n_d$ &    1.343&   1.933&   2.852&   3.698&   4.611&   5.676\\
 \hline\hline
\end{tabular}
\end{center}
\end{table}

In Table~\ref{tab1}, we show results from average-atom calculations of plasmas of
elements in row four of the periodic table. In each case, we assume that the plasma density
is the metallic density and that the plasma temperature is $T=5$~eV. In the first four rows
of the table, we list the atomic symbol, atomic number $Z$, atomic weight $A$ and density $\rho$~(g/cc).
In the next two lines, we show average-atom eigenvalues $\epsilon_{3s}$ and $\epsilon_{3p}$ of
$3s$ and $3p$ states (eV). The theoretical inner shell ionization thresholds -$\epsilon_{nl}$
are 20-30\% smaller than measured values~\cite{PRD}.
Rows seven and eight list $N_b$ and $N_c$, the average number of bound and continuum electrons
inside the Wigner-Seitz (WS) sphere. The
average-atom model predicts that there are approximately 18 bound electrons and
$Z-18$ continuum electrons for each element. The ninth row gives $Z_i$, the average ionic charge or,
equivalently, the number of free electrons per ion. For elements listed in the table,
$Z_i$, which is near 2, is significantly smaller than  $N_c$ owing to the fact that the continuum charge
is concentrated inside the WS sphere. In the final three rows of the table, we list
$n_s$, $n_p$ and $n_d$, contributions to $N_c$ from $s$, $p$ and $d$ partial waves, respectively.
It should be noted that these three partial-wave contributions account almost completely
for $N_c$.  Moreover, for the elements in question, $d$-states provide the dominant
contribution to the sum. In summary, the continuum is concentrated inside the WS
sphere and is dominated by $d$-state partial waves.

\begin{figure}
\centerline{\includegraphics[scale=0.6]{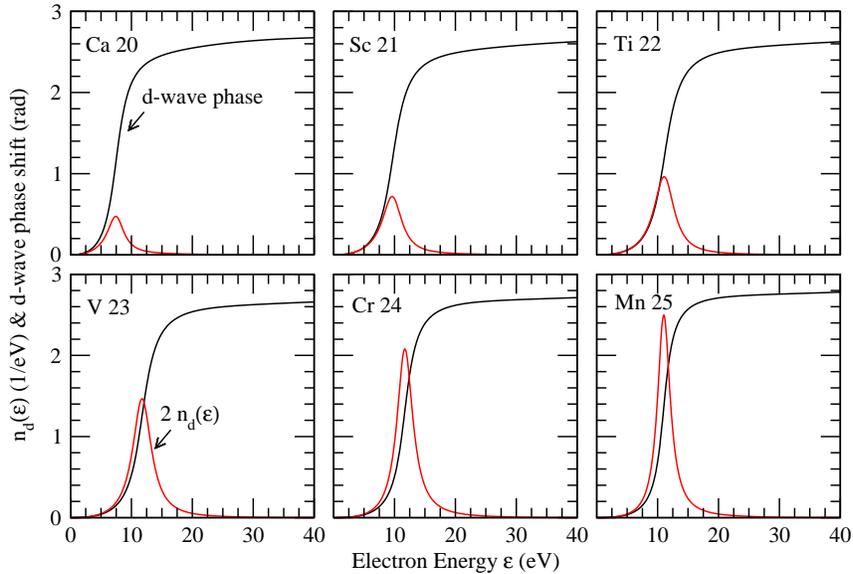}}
\caption{The $d$-state phase shift $\delta_d(\epsilon)$ and twice the
continuum $d$-state distribution function $n_d(\epsilon)$ are shown as functions
of electron energy $\epsilon$~(eV) for
5~eV plasmas at metallic density of
elements in row four of the periodic table.\label{fig1}}
\end{figure}

In Fig.~\ref{fig1}, we plot the $d$-state phase shift $\delta_d(\epsilon)$ and the
continuum $d$-wave distribution
inside the WS sphere
$n_d(\epsilon)$ for elements in Table~\ref{tab1}.
The distribution function $n_l(\epsilon)$ for a continuum state with angular momentum $l$ is defined by
\begin{equation}
 n_l(\epsilon) =
 \frac{2(2 l+1)}{1+\exp{[(\epsilon-\mu)/kT]}} \int_0^{R_{\text{\tiny WS}}}\hspace{-1em}
 P_{p l}(r)^2 dr ,
\end{equation}
where $P_{pl}(r)$ is the radial wave function for a continuum state with momentum $p$ and angular momentum $l$.
When integrated over energy,
\begin{equation}
n_l = \int_0^\infty \!\! n_l(\epsilon)\, d\epsilon
\end{equation}
gives the total number of continuum electrons  for the $s,\, p,\, d$
partial waves listed in the final three rows of Table~\ref{tab1}.
In all cases, the continuum $d$-waves exhibit resonant
behavior near 12~eV, with strong peaks in the distribution functions $n_l(\epsilon)$
and corresponding changes in the phase shifts by almost a factor of $\pi$.

The contribution to the Thomson scattering dynamic structure function
from bound states $S_b(k,\omega)$ is the sum over contributions from individual subshells
with quantum numbers $(n,l)$:
\begin{align}
S_b(k,\omega)    &=\ \sum_{nl} S_{nl}(k,\omega) \\
S_{nl}(k,\omega) &=\ \frac{o_{nl}}{2l+1} \sum_{m} \int \frac{p\, d\Omega_p}{(2\pi)^3}\,
\left| \int\!\! d^3 r\, \psi^\dagger_{\bm p}(\bm{r})\, e^{i\bm{k}\cdot\bm{r}}\,
\psi_{nlm}(\bm{r})
\right|^2 ,\label{snl}
\end{align}
 where $p = \sqrt{2(\omega+\epsilon_{nl})}$
is the momentum of the continuum electron,
$o_{nl}$ is the fractional occupation number of subshell $(n,l)$,
and ${\bm k}={\bm k}_0 - {\bm k}_1$ and $\omega = \omega_0-\omega_1$
are the momentum and energy transfer
from the incident photon $(\bm{k}_0,~\omega_0)$
to the scattered photon $(\bm{k}_1,~\omega_1)$, respectively.
Atomic units in which $e = \hbar = m = 1$ are used here.
In Eq.~(\ref{snl}), $\psi_{\bm p}(\bm{r})$ is an average-atom wave function that
approaches a plane wave $e^{i\bm {p\cdot r}}$ asymptotically and $\psi_{nlm}(\bm{r})$
is the wave function for a bound state with quantum numbers $(n,l,m)$.
The bound-state wave function is expressed in terms of the radial wave function $P_{nl}(r)$ as
\begin{equation}
\psi_{nlm}(\bm{r}) = \frac{1}{r} P_{nl}(r) Y_{lm}(\hat{r}) .
\end{equation}
Moreover, the scattering wave function, which
consists of a plane wave plus an {\it incoming}
spherical wave is expanded as
\begin{equation}
\psi_{\bm{p}}(\bm{r}) = \frac{4\pi}{p} \sum_{l_1m_1} i^{\, l_1}
e^{-i\delta_{l_1}} \frac{1}{r} P_{p l_1}(r)\,
Y^\ast_{l_1 m_1}(\hat{p})\, Y_{l_1 m_1}(\hat{r}),
\end{equation}
where the radial function $P_{p l}(r)$ is normalized to a phase-shifted
sine wave asymptotically:
\begin{equation}
P_{p l}(r) \to \sin(pr-l\pi/2 +\delta_l).
\end{equation}
With the above definitions in mind, Eq.~(\ref{snl}) can be expressed as
\begin{equation}
S_{nl}(k,\omega) = \frac{2p}{\pi}  o_{nl} \sum_{l_1l_2} A_{l_1 l\,l_2}
\left|I_{l_1 l\,l_2}(p,k)\right|^2 , \label{lsum}
\end{equation}
where
\begin{equation}
I_{l_1 l\,l_2}(p,k) = \frac{1}{p}\, e^{i\delta_{l_1}\!(p)}
\!\int_0^{R_\text{\tiny WS}} \hspace{-1em} dr P_{p l_1}(r) P_{nl}(r) j_{l_2}(kr)
\end{equation}
and
\begin{equation}
A_{l_1 l\, l_2} = (2l_1+1)(2l_2+1) \left( \begin{array}{ccc}
l_1 & l & l_2 \\
0 & 0 & 0
\end{array} \right)^{\! 2} .
\end{equation}

As shown earlier, the dominant contribution to the continuum wave function $P_{nl_1}(r)$
is from $d$ waves ($l_1=2$) and
occurs near continuum energy $\epsilon_p = p^2/2 \approx 12$~eV above threshold.  It follows that the dominant
contribution to $S_{3p}(k,\omega)$ is from the two terms with $(l_1, l, l_2) = (2,1,1)$ and
$(2,1,3)$, while
the dominant contribution to $S_{3s}(k,\omega)$ is from the single term
$(l_1, l, l_2) = (2,0,2)$.
Energy conservation implies that the resonant contribution to the scattered photon spectrum
from bound state $nl$ occurs at scattered X-ray energy $\omega_1 = \omega_0 + \epsilon_{nl} -\epsilon$, with
$\epsilon\approx 12$~eV.

\begin{figure}
\centerline{\includegraphics[scale=0.7]{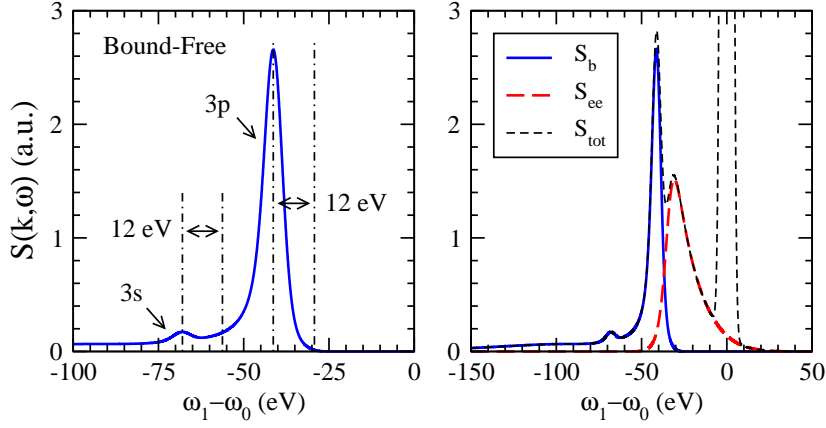}}
\caption{Left panel: $S_b(k,\omega)$ for Cr at metallic density and $T=5$~eV is plotted against
$\omega_1-\omega_0$, where
$\omega_1$ is the scattered photon energy. The incident photon energy is $\omega_0=4750$~eV and the
scattering angle is 40$^\circ$. Resonance peaks and thresholds associated with transitions from $3s$ and $3p$ bound states
are shown by the vertical construction lines.
Right panel: Contributions from bound-free $S_b$, free-free $S_{ee}$, and the total structure function $S_{\text{tot}}$, which
includes elastic scattering, are plotted against $\omega_1-\omega_0$. \label{fig2}}
\end{figure}

As a specific example, we consider scattering of a 4750~eV photon at 40$^\circ$ from Cr at
metallic density and $T=5$~eV. In the left panel of Fig.~\ref{fig2}, we show the bound-free contribution $S_b(k,\omega)$
to the dynamic structure function. The vertical construction lines show the $2s$ and $2p$ resonance peaks and thresholds.
The resonance peaks are shifted 12~eV below
the thresholds of 29.3 and 56.2~eV for the $3s$ and $3p$ electrons, respectively,
 as expected from the position of the peak in $n_d(\epsilon)$
shown in the Cr panel of Fig.~\ref{fig1}.

\begin{figure}
\centerline{\includegraphics[scale=0.6]{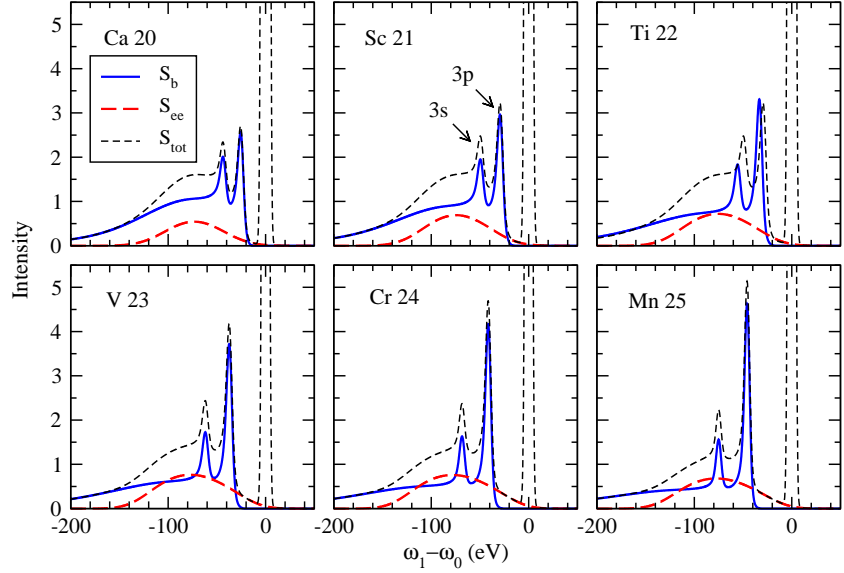}}
\caption{The dynamic structure function $S_\text{tot}(k,\omega)$ (short black dashes), bound-free contributions including $3s$ and
$3p$ resonances $S_b(k,\omega)$ (solid blue lines) and free contributions $S_{ee}(k,\omega)$ (long red dashes) for scattering
from elements listed in Table~\ref{tab1} are plotted as functions of $\omega_1-\omega_0$, where $\omega_1$ is the scattered
X-ray energy and $\omega_0=4750$~eV is the incident
 X-ray energy. The scattering angle for the illustrated cases is  130$^\circ$.
\label{fig3}}
\end{figure}

In Fig.~\ref{fig3}, we present results of calculations of
the inelastic bound-free contribution to the dynamic structure function $S_b(k,\omega)$,
the inelastic free-free contribution $S_{ee}(k,\omega)$ and the total dynamic structure function
including elastic scattering $S_{\text tot}(k,\omega)$
for elements listed in Table~\ref{tab1}. The incident X-ray energy in these
examples is $\omega_0 = 4750$~eV and the scattering angle is
130$^\circ$. The relative amplitude of the resonant peaks is seen to increase systematically throughout the sequence
as expected from the $d$-electron distributions shown in Fig.~\ref{fig1}. By contrast, the relative amplitude of the free-free
contribution remains relatively constant from element to element as a consequence of the fact that $Z_i$, the number of free
electrons per ion, is relatively constant throughout the sequence.

In conclusion, the resonances predicted in the bound-free contribution to the Thomson scattering X-ray spectrum
arise from a $d$-state resonance in the continuum associated with continuum lowering in a dense plasma.
For the examples considered herein, elements between Ca and Mn at metallic density and $T=5$~eV, the $d$-state resonances
occur at continuum-electron energy $\epsilon\approx 12$~eV. The amplitude of the $d$-state resonances increase systematically
from element to element. The $d$-state resonance dominate the continuum and the bound-free contribution to the
 dynamic-structure function $S_b(k,\omega)$. The resonance associated with transitions from subshell $nl$ occurs
at scattered X-ray energy $\omega_1 = \omega_0 +\epsilon_{nl}-\epsilon$, where the resonance energy
for the cases considered here is $\epsilon \approx 12$~eV.

\section*{Acknowledgements}
The authors owe a debt of gratitude to T. D\"{o}ppner for bringing up the question of the
origin of the resonances predicted in average-atom calculations.
The work of J.N.\ and K.T.C.\ was performed under the auspices of the U.S.\
Department of Energy by Lawrence Livermore National Laboratory
under Contract DE-AC52-07NA27344.






\end{document}